\documentclass[showpacs,aps,prd,nofootinbib,floatfix,amsmath,amssymb,superscriptaddress]{revtex4}
\usepackage{graphicx}
\makeatletter
\newbox\slashbox \setbox\slashbox=\hbox{$/$}
\newbox\Slashbox \setbox\Slashbox=\hbox{\large$/$}
\def\pFMslash#1{\setbox\@tempboxa=\hbox{$#1$}
  \@tempdima=0.5\wd\slashbox \advance\@tempdima 0.5\wd\@tempboxa
  \copy\slashbox \kern-\@tempdima \box\@tempboxa}
\def\pFMSlash#1{\setbox\@tempboxa=\hbox{$#1$}
  \@tempdima=0.5\wd\Slashbox \advance\@tempdima 0.5\wd\@tempboxa
  \copy\Slashbox \kern-\@tempdima \box\@tempboxa}

\def\miss#1{\ifmmode{/\mkern-11mu #1}\else{${/\mkern-11mu #1}$}\fi}
\makeatother

\begin{document}

\title{Rare three-body decay $t \to c h \gamma$ in the standard
model and the two-Higgs doublet model}

\date{\today}
\author{A. Cordero-Cid}
\affiliation{Facultad de Ciencias F\'\i sico Matem\'aticas,
Benem\'erita Universidad Aut\'onoma de Puebla, Apartado Postal 1152,
Puebla, Pue., M\' exico}
\author{J. L. Garc\'{\i}a-Luna}
\affiliation{ Departamento de F\'{\i}sica,  Centro Universitario de
Ciencias Exactas e Ingenier\'{\i}as, Universidad de Guadalajara,
Blvd. Marcelino Garc\'{\i}a Barrag\'an 1508, C.P. 44840, Guadalajara
Jal., M\'exico}
\author{F. Ram\'irez-Zavaleta}
\affiliation{Departamento de F\'{\i}sica, CINVESTAV, Apartado Postal
14--740, 07000, M\'exico D. F., M\'exico}
\author{G. Tavares-Velasco}
\affiliation{Facultad de Ciencias F\'\i sico Matem\'aticas,
Benem\'erita Universidad Aut\'onoma de Puebla, Apartado Postal 1152,
Puebla, Pue., M\' exico}
\author{J. J. Toscano}
\affiliation{Facultad de Ciencias F\'\i sico Matem\'aticas,
Benem\'erita Universidad Aut\'onoma de Puebla, Apartado Postal 1152,
Puebla, Pue., M\' exico}

\begin{abstract}

A complete calculation of the rare three-body decay $t \to c h
\gamma$ is presented in the framework of the standard model. In the
unitary gauge, such a calculation involves about 20 Feynman
diagrams. We also calculate this decay in the general two-Higgs
doublet model (model III), in which it arises at the tree-level.
While in the standard model the decay $t \to c h \gamma$ is
extremely suppressed, with a branching fraction of the order of
$10^{-15}$ for a Higgs boson mass of the order of 115 GeV, in the
model III it may have a branching ratio up to $10^{-5}$. We also
discuss the crossed decay $h \to b \bar{s} \gamma$.
\end{abstract}


\maketitle
\section{introduction}

Although the standard model (SM) has been tested to a great
accuracy, it is worth investigating some rare processes as they may
represent a detailed test for this theory in the current and future
particle accelerators. Among these processes,  top quark decays have
attracted considerable attention due in part to the extraordinary
disparity between the top quark mass and those of the remaining
quarks, which suggests that the former may give rise to the
appearance of new phenomena \cite{Chakraborty:2003iw}. For instance,
it has been conjectured that the top quark  may play an important
role in the mechanism of electroweak symmetry breaking
\cite{topcolor}. The interest in top quark physics also stems from
the advent of the CERN large hadron collider (LHC), which will allow
the copious production of about $10^7$-$10^8$ top quark pairs per
year. This will be useful to examine to a high accuracy several top
quark properties, such as decay channels other than the main one
$t\to bW$. Due to the large top quark mass, it can have a wide
spectrum of decay modes. In fact the top quark is likely to be the
only SM particle to decay into a Higgs boson plus one or more other
particles. In the SM, even the second most likely decay modes, the
nondiagonal ones $t\to sW$ and $t\to dW$, have very small branching
ratios, of the order of $10^{-3}$-$10^{-4}$
\cite{Chakraborty:2003iw}. The top quark decay $t\to bWZ$ has a tiny
branching ratio, but it was believed
\cite{Mahlon:1994us,Decker:1992wz,Altarelli:2000nt} it might be
useful to probe the top quark mass due to the fact that this decay
mode is close to the kinematical threshold. For a Higgs boson mass
of the order of $120$ GeV, the branching ratio for the decay channel
$t\to bWH$ is about $10^{-8}$ \cite{Decker:1992wz}. Another
three-body decay, $t\to cWW$, is much more suppressed by the
Glashow-Illiopoulus-Maiani (GIM) mechanism: its branching ratio is
of the order of $10^{-13}$ \cite{Jenkins:1996zd}. One-loop induced
flavor changing neutral current (FCNC) decays of the top quark seem
to be far from the reach of detection, though they can have sizeable
branching ratios in some extended theories. In fact, the search for
large signatures of FCNCs involving the top quark is considered the
ultimate test for the SM \cite{Han:1995pk}. The following FCNC top
quark decays have been widely studied in the SM and some of its
extensions: $t\to ch$
\cite{Eilam:1990zc,Mele:1999zk,Hou:1991un,Yang:1993rb,Eilam:2001dh,Guasch:1999jp,Bejar:2000ub},
$t \to c V$ ($V=\gamma,\,g,\,Z$)
\cite{Eilam:1990zc,Diaz1,Li:1993mg,Couture:1994rr,
Yang:1997dk,Lopez:1997xv,deDivitiis:1997sh,Yue:2001qr,Lu:2003yr}, $t
\to c V_i V_j$ \cite{Diaz-Cruz:1999ab}, and more recently other rare
decays \cite{Cordero-Cid:2004hk}. While the decay modes $t \to c
V_i$ and $t\to ch$ all have branching ratios below the $10^{-10}$
level in the SM \cite{Eilam:1990zc,Mele:1999zk}, they can be
dramatically enhanced beyond the SM. For instance, in the minimal
supersymmetric standard model (MSSM) with broken $R$ parity the
upper limits are  \cite{deDivitiis:1997sh}: $Br(t\to c\gamma)\sim
10^{-5}$, $Br(t\to cg)\sim 10^{-3}$, $Br(t\to c Z)\sim 10^{-3}$, and
$Br(t\to ch)\sim 10^{-4}$. Two-Higgs doublet models (THDMs) can also
give rise to large enhancements for this class of decays. In
particular, the decay $t\to ch$ may have a branching ratio up to $
10^{-2}$ in the THDM of type III \cite{Hou:1991un}.

The aim of this work is to discuss the decay $t \to c h \gamma$ and
the crossed one $h \to \bar{q_i}q_j \gamma$. These FCNC decay modes
are interesting since they involve the Higgs boson, which still
remains the most elusive piece of the SM. Since these processes are
expected to be strongly suppressed by the GIM mechanism, which
effectively suppresses  FCNC transitions involving virtual down-type
quarks, they are very sensitive to any new physics effects. The
$t\to ch\gamma$ decay occurs at the one-loop level in the SM. In the
unitary gauge there are about 20 Feynman diagrams. For completeness,
we will present explicit results for this calculation. On the other
hand, as already mentioned, some SM extensions may give rise to
large FCNC effects. In this context, we will consider the specific
case of the general two-Higgs doublet model type III
\cite{Cheng:1987rs,Antaramian:1992ya,Hall:1993ca,Luke:1993cy}, which
allows for tree-level FCNCs, unlike the type-I and type-II THDMs,
where FCNCs are removed by invoking an {\it ad hoc} symmetry
\cite{Glashow:1976nt}. We will show below that this model may
enhance considerably the decay $t\to ch\gamma$ due in part to the
tree-level FCNCs.

The rest of the paper is organized as follows. In Sec. II we discuss
the most important details of the $t\to ch\gamma$ calculation within
the SM. Although the formulas for the decay $t\to ch\gamma$  are too
lengthy, they are presented in Appendix A for completeness. The
scenario that arises in the THDM is discussed in Sec. III. Finally,
the conclusions are presented in Sec. IV.

\section{Decay $t\to ch\gamma$ in the SM}

\begin{figure}
\centering
\includegraphics[width=3in]{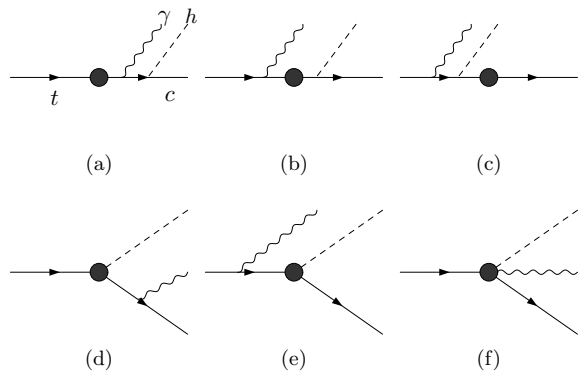}
\caption{\label{diag5} Feynman diagrams contributing to the decay
$t\to ch\gamma$. An extra set of diagrams is obtained when the Higgs
boson and the photon are exchanged. The blob represents the one-loop
contributions from irreducible Feynman diagrams. In the massless
charm quark limit, the diagrams in which the Higgs boson emerges
from the $c$ quark give no contribution as the coupling $h \bar{c}c$
is proportional to $m_c$.}
\end{figure}

\begin{figure}
\centering
\includegraphics[width=2in]{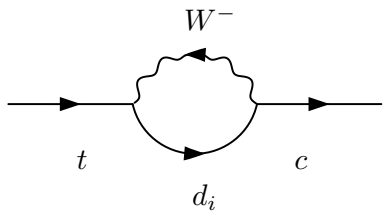}
\caption{\label{diag1} One-loop contribution to the $tc$ vertex in
the unitary gauge. ${d_i}$ stands for a generic down quark.}
\end{figure}

\begin{figure}
\centering
\includegraphics[width=3in]{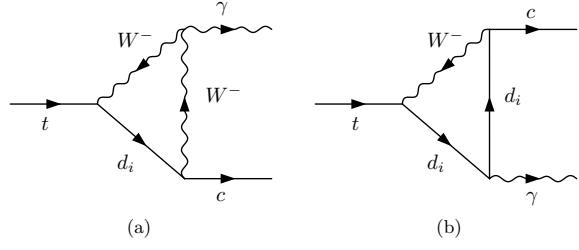}
\caption{\label{diag3} Irreducible Feynman diagrams contributing to
the  $tc\gamma$ vertex in the unitary gauge.}
\end{figure}

\begin{figure}
\centering
\includegraphics[width=3in]{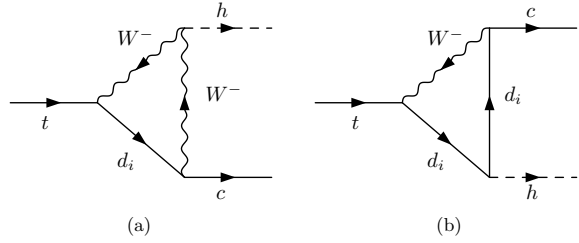}
\caption{\label{diag2} The same as in Fig. \ref{diag3} for the $tch$
vertex.}
\end{figure}

\begin{figure}
\centering
\includegraphics[width=3.5in]{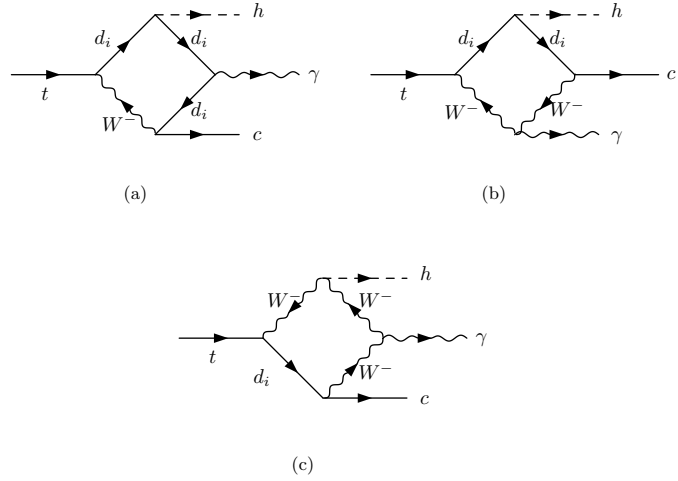}
\caption{\label{diag4} The same as in Fig. \ref{diag3} for the
$tch\gamma$ vertex. There is also another set of box diagrams where
the Higgs boson and the photon are exchanged.}
\end{figure}

We turn to the most relevant details of the calculation of the decay
$t\to ch\gamma$ in the SM. In the unitary gauge, this decay proceeds
through 20 Feynman diagrams, which are depicted in Fig. \ref{diag5},
where the blob represents one-loop contributions. There are
contributions from loops carrying charged $W$ gauge bosons and down
quarks, which we will denote generically by $d_i$. We have grouped
these diagrams into four sets: those that arise from the irreducible
vertices $tc$, $tch$, and $tc\gamma$, as well as the box diagrams.
The loops are shown explicitly through Fig. \ref{diag1} to Fig.
\ref{diag4}. We used the Passarino-Veltman reduction scheme
\cite{Passarino:1978jh} to calculate the amplitudes for each set of
diagrams. As a check of the calculation we have verified explicitly
the cancelation of ultraviolet singularities and fulfillment of
electromagnetic gauge invariance. It turns out that the amplitude
for the set of diagrams arising from the $tc$ vertex is ultraviolet
divergent, but the divergences are exactly canceled by those
appearing in the set of diagrams arising from the $tc\gamma$ vertex.
Furthermore, the amplitudes of these two sets of diagrams should be
combined to give a gauge invariant amplitude. As far as the
remaining contributions are concerned, although the set of diagrams
arising from the $tch$ vertex yields an ultraviolet finite amplitude
by its own, and so does the set of box diagrams, gauge invariance is
only achieved when these two amplitudes are added together. It is
interesting to note that these properties verify only if those terms
that are independent of the internal down quark mass $m_{d_i}$ are
dropped. Those terms cancel when one sums over the three quark
families since by unitarity of the CKM matrix $\sum_{d_i}
V_{td_i}V^\dagger_{d_i c}=0$, which is the GIM mechanism.

We now proceed to discuss in more detail the analytical results. We
will denote the 4-momenta of the participating particles as follows

\begin{equation}
t(p_1)\to c(p_2) h(q)\gamma(k),
\end{equation}
$\alpha$ stands for the photon momentum Lorentz index, and we
introduce the scaled variables $x$, $y$ and $z$, given by $k\cdot
p_1=m_t^2x/2$, $p_1\cdot q=m_t^2y/2$, and $p_1\cdot p_2=m_t^2z/2$,
along with $\mu_h=m_h^2/m_t^2$. From 4-momentum conservation it
follows that $z=2-x-y$. In the rest frame of the decaying $t$ quark,
$x$, $y$ and $z$ are related to the energies of the final particles
as follows: $x=2\,E_\gamma/m_t$, $y=2\,E_h/m_t$, and $z=2\,E_c/m_t$.

Before presenting the results, it is worth discussing the gauge
invariance of the transition amplitude under $U(1)_{em}$. The
calculation of the Feynman diagrams via the Passarino-Veltman
reduction scheme leads to the following expression for the
transition amplitude

\begin{equation}
\label{amplitude} {\cal M}(t\to c
h\gamma)=\epsilon^*_\alpha(k)\cdot\,\bar{u}_c(p_2)\left({\cal
M}^\alpha_L P_L+{\cal M}^\alpha_R P_R\right)u_t(p_1),
\end{equation}
where $P_L=(1-\gamma^5)/2$, $P_R=(1+\gamma^5)/2$,
\begin{equation}
{\cal M}^\alpha_{L}={A_1}_{L}p_1^\alpha
+{A_2}_{L}p_2^\alpha+{A_3}_{L}\,\sigma^{\alpha\mu}k_\mu+{A_4}_{L}p_1^\alpha
\pFMSlash{k}+{A_5}_{L}\gamma^\alpha+{A_6}_{L}p_2^\alpha
\pFMSlash{k},
\end{equation}
and a similar expression for ${\cal M}^\alpha_R$. The coefficients
${A_i}_{L,R}$  include the contributions of the three quarks $d_i$
that circulate in the loops.  For this amplitude to be gauge
invariant under $U(1)_{em}$, it must vanish when the polarization
vector of the photon $\epsilon^*_\alpha(k)$ is replaced by its
four-momentum $k_\alpha$, i.e.,  $k_\alpha({\cal M}_L^\alpha
P_L+{\cal M}_R^\alpha P_R) =0$, which is the Ward identity. One way
to achieve this is that each term of the left side of the identity
vanishes separately, which is equivalent to

\begin{equation}
(k\cdot p_1){A_1}_{L,R} +(k \cdot
p_2){A_2}_{L,R}+{A_3}_{L,R}\,\sigma^{\alpha\mu}k_\mu
k_\alpha+\left((k\cdot p_1){A_4}_{L,R} +{A_5}_{L,R}+(k\cdot
p_2){A_6}_{L,R}\right) \pFMSlash{k}=0. \footnote{The subindex $L,R$
means that the identity is true for either ${A_i}_L$ or ${A_i}_R$.}
\end{equation}
Since $\sigma^{\alpha\mu}k_\mu k_\alpha=0$,  only the ${A_3}_{L,R}$
term vanishes automatically. It means that the remaining
coefficients ${A_i}_{L,R}$ are not independent since the only way to
fulfill the above equation is that $(k\cdot p_1){A_1}_{L,R} +(k
\cdot p_2){A_2}_{L,R}=0$ and $(k\cdot p_1){A_4}_{L,R}
+{A_5}_{L,R}+(k\cdot p_2){A_6}_{L,R} =0$. Therefore, if we showed
that the coefficients obtained in our calculation fulfill these
relations then we would have shown that the transition amplitude is
gauge invariant under $U(1)_{em}$. We have verified explicitly that
the amplitude obtained from our calculation obeys these relations.
We can thus express ${A_2}_{L,R}$ in terms of ${A_1}_{L,R}$, whereas
${A_5}_{L,R}$ can be expressed in terms of ${A_4}_{L,R}$ and
${A_6}_{L,R}$. Using these results, the ${\cal M}_{L,R}$ amplitude
can be rewritten in the form

\begin{eqnarray}
\label{amplitude-1} {\cal
M}^\alpha_{L,R}&=&\frac{{F_1}_{L,R}}{m_t}\left((k\cdot
p_2)p_1^\alpha -(k\cdot
p_1)p_2^\alpha\right)+i\,m_t\,{F_2}_{L,R}\,\sigma^{\alpha\mu}k_\mu+{F_3}_{L,R}\left(p_1^\alpha
\pFMSlash{k}-(k\cdot
p_1)\gamma^\alpha\right)\nonumber\\&+&{F_4}_{L,R}\left(p_2^\alpha
\pFMSlash{k}-(k\cdot p_2)\gamma^\alpha\right).
\end{eqnarray}
where the new coefficients ${F_i}_{L,R}$ are given in terms of the
old coefficients ${A_i}_{L,R}$. It is easy to see that the above
equation vanishes when  contracted with $k_\alpha$. The coefficients
${F_i}_{L,R}$ are too cumbersome to be presented here, we will
content with presenting the results in the limit of a massless charm
quark, in which case the charm quark becomes purely left-handed,
namely, $u_c(p_2)\to P_L u_c(p_2)$ or $\bar{u}_c(p_2)\to
\bar{u}_c(p_2) P_R $ . As a result, the ${F_1}_{L}$, ${F_2}_{L}$,
${F_3}_{R}$, and ${F_4}_{R}$ terms must vanish in this limit. We
have also verified that this is true by setting $m_c=0$ in the
general results. On the other hand, we cannot set $m_{d_i}=0$ since
the whole amplitude would vanish when summing over the three quark
families due to the GIM mechanism.

After squaring the amplitude (\ref{amplitude}) we average over
initial spins and sum over final polarizations to obtain, in the
$m_c= 0$ limit:
\begin{eqnarray}
\label{mcuad} |{\cal M}(t\to c
h\gamma)|^2&=&\frac{m_t^6}{8}\Big[u\,z(x\,z-\,u)|{F_1}_R|^2+8\,x\,u\,|{F_2}_R|^2
+2\,x^2\,z|{F_3}_L|^2+2\,u^2\,z|{F_4}_L|^2+4\,u\,{\rm
Re}\Big(2\,x\,{F_3}_L{F_2}_R^\dag\nonumber\\
&+&(x\,z-u)\big({F_1}_R{F_2}_R^\dag+{F_3}_L{F_1}_R^\dag\big)
+u\big({F_3}_L{F_4}_L^\dag+2\,{F_4}_L{F_2}_R^\dag\big)\Big)\Big],
\end{eqnarray}
where we introduced the auxiliary variable  $u=1+\mu_h-y$.

\subsection{Decay width}

\begin{figure}
\centering
\includegraphics[width=3.5in]{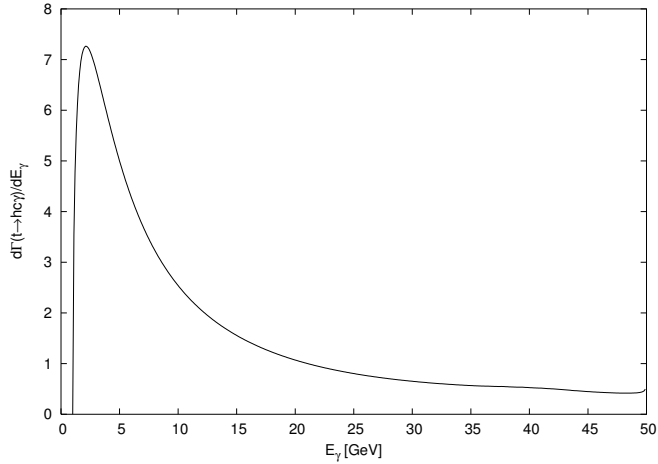}
\caption{\label{enerdist} Photon energy distribution for the decay
$t\to ch\gamma$ in the SM and for $m_h=115$ GeV. The vertical scale
is in units of $10^{-17}$. We have imposed a cutoff of $E_\gamma>1$
GeV to tame the infrared singularity.}
\end{figure}

From the square amplitude,  we obtain the photon energy
distribution, which is given by
\begin{equation}
\label{photonenergy} \frac{d\Gamma(t\to c
h\gamma)}{dx}=\int_{1-x+\frac{\mu_h}{1-x}}^{1+\mu_h}|{\cal M}(t\to c
h\gamma)|^2dy,
\end{equation}
whereas the decay width reads
\begin{equation}
\label{decay_width} \Gamma(t\to c
h\gamma)=\frac{m_t}{256\pi^3}\int_{x_{min}}^{1-\mu_h}\frac{d\Gamma(t\to
c h\gamma)}{dx} dx,
\end{equation}
where $x_{min}=2{E_\gamma}_{min}/m_t$, with ${E_\gamma}_{min}$ being
an arbitrary minimum value for the photon energy. We cannot
integrate over the whole photon energy spectrum since the
denominator of the amplitude coming from the Feynman diagrams where
the photon emerges from the external top quark have a factor
$(p_1-k)^2-m_t^2=-2k\cdot p_1=-2m_t E_\gamma$, which vanishes when
$E_\gamma=0$. This is an infrared singularity which reflects the
fact that a zero energy photon cannot be experimentally detected.
The infrared nature of the transition amplitude can be observed in
Fig. \ref{enerdist}, where we have plotted the $t\to c h\gamma$
photon energy distribution for $m_h=115$ GeV.

\begin{figure}[!hbt]
\centering
\includegraphics[width=3.5in]{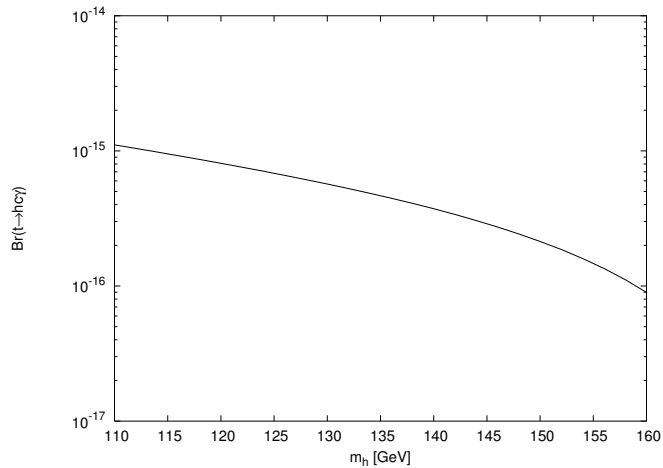}
\caption{\label{brsm} $Br(t\to ch\gamma)$ in the SM as a function of
the Higgs boson mass. We consider $E_\gamma>1 $ GeV.}
\end{figure}

The branching fraction follows easily after dividing
(\ref{decay_width}) by the main top quark decay width $\Gamma(t\to
bW)$. Using the current values for the SM parameters
\cite{Eidelman:2004wy}, numerical integration of Eq.
(\ref{decay_width}) gives the result $Br(t\to c h\gamma)\sim
10^{-15}$ GeV for a Higgs boson mass around 115 GeV and
${E_\gamma}_{min}=1$ GeV. For a heavier Higgs boson the branching
ratio is one order below, as shown in Fig. \ref{brsm}. In obtaining
these numerical results, the Passarino-Veltman scalar form factors
were evaluated numerically via the FF routines
\cite{vanOldenborgh:1990yc}. This very suppressed result is mainly
due to the GIM mechanism and phase space suppression. It is somewhat
interesting to assess how each single term in Eq. (\ref{mcuad})
contributes to the decay width. In Table \ref{resultados-sm} we
present the partial contribution of each term appearing in Eq.
(\ref{mcuad}) for $m_h=115$ GeV. We see that the largest
contribution comes from the coefficient ${F_1}_R$, whereas the
coefficient ${F_4}_L$ gives a contribution one order of magnitude
below.

\begin{table}[!htb]
\caption{\label{resultados-sm}Partial contribution  to the $Br(t\to
ch\gamma)$ from each term in Eq. (\ref{mcuad}) for $m_h=115$ GeV.}
\begin{tabular}{cccccccccc}
\hline &$|{F_3}_L|^2$&$|{F_4}_L|^2$&$|{F_1}_R|^2$&$|{F_2}_R|^2$&
${F_3}_L{F_2}_R^\dag$&${F_1}_R{F_2}_R^\dag$&${F_3}_L{F_1}_R^\dag$&${F_3}_L{F_4}_L^\dag$&${F_4}_L{F_2}_R^\dag$\\
\hline
Contribution$\times 10^{16}$&2.2&0.13&9.7&5.4&-4.8&2.4&-0.24&0.4&-0.9\\
\hline
\end{tabular}
\end{table}

\section{Decay $t\to ch\gamma$ in the THDM-III}
In the THDM-III, the quarks are allowed to couple simultaneously to
more than one scalar doublet \cite{Cheng:1987rs}. This leaves open
the possibility of sizeable effects in the scalar FCNC couplings
involving quarks of the second and third generations. Unlike the
first and second versions of the THDM, in model III no {\it{ad hoc}}
symmetries are invoked to eliminate tree-level scalar FCNC couplings
but instead a more realistic pattern for the Yukawa matrices is
imposed and constraints on the scalar FCNC are derived from
phenomenology \cite{Atwood:1996vj}.  The tree-level scalar FCNC
interactions are given by

\begin{equation}
\label{lag} {\cal{L}}_{Y,FCNC}^{III} = \xi_{ij} \sin{\alpha}
\bar{f_i} f_j h + \xi_{ij} \cos{ \alpha} \bar{f_i} f_j H + \xi_{ij}
\cos{\alpha} \bar{f_i} \gamma^5 f_j A + {\rm H.c.},
\end{equation}

\noindent where we are using the Higgs mass-eigenstate basis with
the light and heavy CP-even Higgs bosons $h$ and $H$, and the CP-odd
Higgs boson $A$, $\alpha$ denotes the mixing angle, and $\xi_{ij}$
corresponds to the off-diagonal Yukawa couplings. It is usual to use
the parametrization introduced by Cheng and Sher in Ref.
\cite{Cheng:1987rs}: $\xi_{ij}= \lambda_{ij} \sqrt{m_i m_j}/v$,
where the mass factor gives the strength of the interaction, whereas
the dimensionless parameters $\lambda_{ij}$ are usually assumed of
order unity. Although the couplings involving light quarks are
naturally suppressed according to this parametrization, the
interaction $tc\phi$, with $\phi$ any of the three physical Higgs
bosons of the THDM, is much less suppressed. Therefore, it is
interesting to examine to what extent the decay $t\to ch\gamma$ can
be enhanced by this model.

\begin{figure}[!hbt]
\centering
\includegraphics[width=3.5in]{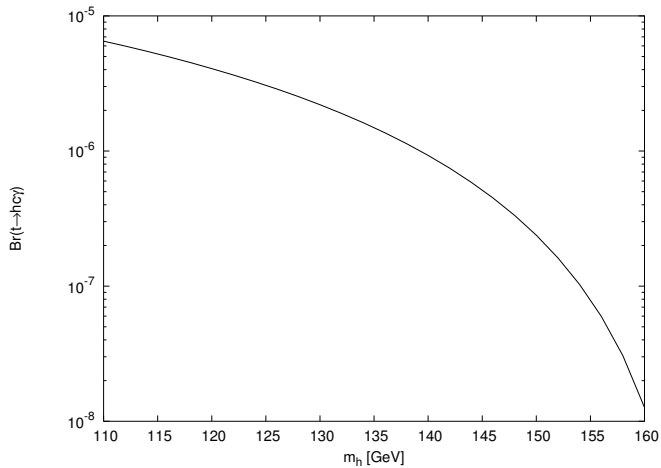}
\caption{\label{brthdm}$t\to ch\gamma$ branching ratio in the
THDM-III as a function of the Higgs boson mass. We assumed
$E_\gamma\ge 10$ GeV and $\lambda_{tc}\sim 1$.}
\end{figure}

The tree-level Feynman diagrams contributing to $t\to c\phi\gamma$
are similar to those shown in Fig. \ref{diag5}(d) and
\ref{diag5}(e). For illustration purposes it is enough to consider
the decay into the lightest CP-even Higgs boson $h$. We will omit
the factor $\sin\alpha$, which is to be reinserted when necessary.
We will calculate the decay rate without neglecting the $c$ quark
mass. The transition amplitude can be arranged as in Eqs.
(\ref{amplitude}) and (\ref{amplitude-1}):

\begin{equation}
\label{ampthdm} {\cal M}^{\rm III}(t\to
ch\gamma)=-\frac{i\pi\,\alpha\,\lambda_{tc}}{3\,s_W\,m_W}\frac{\sqrt{m_t\,m_c}}{
k\cdot p_1\,k\cdot p_2}{\bar u}_c(p_2)\Big((k\cdot q)\,
i\,\sigma^{\alpha\mu}k_\mu+2\left((k\cdot p_2)p_1^\alpha -(k\cdot
p_1)p_2^\alpha\right)\Big)u_t(p_1)\cdot \epsilon^*_{\alpha},
\end{equation}
As discussed above, this amplitude vanishes when
$\epsilon^*_{\alpha}$ is replaced by $k_\alpha$, thereby being
$U(1)_{em}$ gauge invariant. The square amplitude reads
\begin{eqnarray}
|{\cal M}^{\rm III}(t\to
ch\gamma)|^2&=&\left(\frac{4\,\pi\,\alpha\,\lambda_{tc}}{3}\right)^2
\frac{\sqrt{\mu_c} m_t^2}{2\,s_W^2\,m_W^2\,u^2\,x^2}\Big(u\,( u^2\,(
2 + x ) +2\,(x+1)(y-2)u + x\,( x^2 + 2\,(y-2)(x+y-2)))\nonumber\\&-&
4\,u\,( u + x\,(x + y-2 ) )\,\mu_c^{1/2} + 2\,x^2\,(u+y-2) \,\mu_c-
4\,x^2\,\mu_c^{3/2}\Big),
\end{eqnarray}
where $\mu_c=m_c^2/m_t^2$ and $u$ is now defined as
$u=1+\mu_h-\mu_c-y$. This result can be inserted into Eq.
(\ref{decay_width}) to obtain the $t\to ch\gamma$ branching
fraction. However, the above result is also infrared divergent and
we should be careful when integrating over the  photon energy.
Assuming and idealized situation, we will calculate the decay width
in the rest frame of the $t$ quark and impose a minimum cut of $10$
GeV on the photon energy. This is equivalent to introduce a
fictitious photon mass $m_\gamma=10$ GeV. The integration limits are
thus

\begin{equation}
2\sqrt{\mu_\gamma}\le x\le 1+\mu_\gamma-\mu_h-\mu_c-2\,\sqrt{\mu_c\,
\mu_h},
\end{equation}

\begin{equation}
y_{min,\,max}=\frac{1}{2(1-x-\mu_\gamma)}\left((2-x)(1+\mu_\gamma+\mu_h-\mu_c-x)
\mp\sqrt{x^2-4\,\mu_\gamma}\,\lambda^{\frac{1}{2}}(1+\mu_\gamma-x,\mu_h,\mu_c)\right),
\end{equation}
with $\mu_\gamma=m_\gamma^2/m_t^2$ and
$\lambda(x,y,z)=x^2+y^2+z^2-2(xy+xz+yz)$. Assuming $\lambda_{tc}\sim
1$, numerical integration of Eq. (\ref{decay_width}) yields $Br(t\to
ch\gamma)\sim 10^{-4}$ for $m_h$ around 115 GeV.  In  Fig.
\ref{brthdm} we have plotted the $t\to ch\gamma$ branching ratio as
a function of $m_h$. For $m_h$ ranging between 110 and 140 GeV,
$Br(t\to ch\gamma)$ is of the order of $10^{-5}$, but it decreases
quickly as $m_h$ approaches the top quark mass.

\subsection{The decay $h\to b\bar{s}\gamma$}
\begin{figure}[!hbt]
\centering
\includegraphics[width=3.5in]{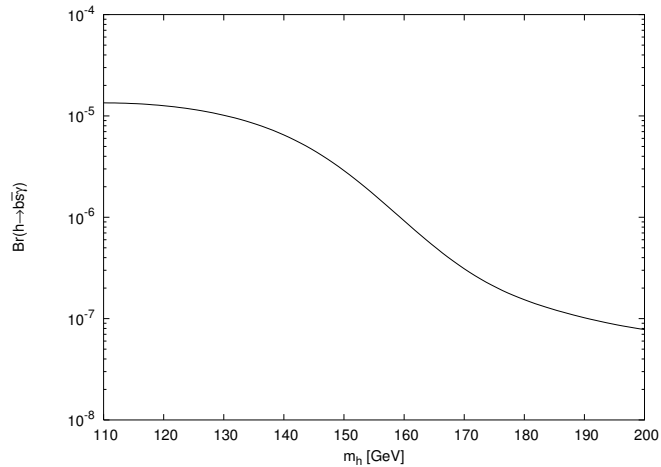}
\caption{\label{brhtobsg} $h\to b\bar{s}\gamma$ branching ratio in
the THDM-III as a function of the Higgs boson mass. We assumed
$E_\gamma\ge 10$ GeV and $\lambda_{bs}\sim 1$.}
\end{figure}

It is also interesting to consider the crossed decay $h\to b\bar{s}
\gamma$, which is also very suppressed in the SM. The two-body decay
$h\to b\bar s$ has already been calculated in the SM \cite{Haeri},
the THDM \cite{Arhrib:2004xu}, and the MSSM \cite{Bejar:2004rz}. It
has been found that this decay mode may be at the reach of future
colliders. In the THDM-II, the two-body decay $h\to b{\bar s}$ as
well as the $h\to b{\bar s}\gamma$ one are suppressed by a factor
$\sqrt{m_b m_s}$, which enters into the Cheng-Sher ansatz for the
$Hb\bar s$ coupling. The numerical calculation yields the $h\to
b{\bar s}\gamma$ branching ratio shown in Fig. \ref{brhtobsg} as a
function of $m_h$. We assumed that the total decay width of the
Higgs boson is approximately the SM one, which was calculated via
the {\small HDECAY} program \cite{Djouadi:1997yw}. For 115 GeV $\le
m_h\le $ 130 GeV the main decay channel of the Higgs boson is $h\to
b\bar b$. Around $m_h=130$ GeV, the channel $h\to WW^*$ becomes more
important, and for $m_h\ge 2\,m_W$ the $h\to WW$ mode, with two $W$
real, becomes the main decay channel. So the $h\to b{\bar s} \gamma$
decay start to decrease dramatically for $m_h$ around 140 GeV.

\section{conclusion}

Although the top quark can have a wide spectrum of decay modes due
to its large mass, it has a  very restrictive dynamical behavior
according to the SM predictions. This means that this particle may
be very sensitive to new physics effects, which is strongly
suggested by several SM extensions, which predict sizeable branching
ratios for some rare top quark decay modes. In this paper we have
presented an explicit calculation of the decay $t\to ch\gamma$  both
in the SM and the THDM-III. As occurs with the FCNC two-body decay
$t \to c h$, the three-body decay $t \to c h \gamma$ is negligibly
small in the SM due to the GIM mechanism and phase space
suppression. The reason why the decay width is so small even if
there are infrared singularities is because the GIM mechanism
strongly suppresses those loops diagrams carrying down quarks. In
contrast, in the THDM-III the decay $t \to c h \gamma$ can be
dramatically enhanced in part due to the existence of tree-level
scalar FCNCs but also because of infrared singularities. In this
model the $t\to ch\gamma$ branching ratio can be up to ten orders of
magnitude larger than in the SM. So it can be an alternative mode to
search for FCNC effects. Notice that in order to tame the infrared
singularities, we integrate the decay width imposing a cut off on
the photon energy. Although we calculate the decay width in the SM
using  a minimum photon energy of 1 GeV, the result is still
strongly suppressed, whereas in the THDM there is a dramatic
enhancement even if we use a cut off of 10 GeV.

As far as the crossed decay $h\to d_i \bar{u_j} \gamma$ is
concerned, this is also very suppressed in the SM.  In the THDM-III,
the $h\to d_i \bar{u_j} \gamma$ branching ratio is proportional to
$m_{d_i}\,m_{u_j}$, so it gets somewhat suppressed for external
light quarks. In particular, $Br(h\to b\bar{s}\gamma)\sim 10^{-5}$
for $m_h\sim 115$ GeV, but it decreases dramatically for a heavier
Higgs boson as more decay channels get opened.

\begin{acknowledgments}
We acknowledge support from SNI and SEP-PROMEP (M\'exico). Partial
support from Conacyt under grant No. U44515-F is also acknowledged.
\end{acknowledgments}

\appendix
\section{Amplitudes for the SM decay $t\to ch\gamma$}

In this appendix we present the amplitudes for the decay $t\to
ch\gamma$ in terms of Passarino-Veltman form factors
\cite{Passarino:1978jh}. We split the total amplitude into five
pieces. The ${F_i}_{L,R}$ coefficients arising from the $tc$ vertex
plus the $tc\gamma$ one will be denoted by the superscript
$tc+tc\gamma$, whereas $tch$ will denote those contributions arising
from the irreducible $tch$ vertex alone. As for the box diagrams,
$\rm Box_1$, $\rm Box_2$, and $\rm Box_3$ will denote the
contributions from the box diagrams with one, two and three internal
$W$ gauge bosons, respectively. As discussed in Sec. II, the
contribution denoted by the superscript $tc+tc\gamma$ is ultraviolet
finite by itself, whereas the remaining contributions, $tch$ and
${\rm Box}_i$, give an ultraviolet finite amplitude by its own, but
they should be added together to obtain gauge invariance. We do not
present below any non gauge invariant terms since they cancel each
other when adding the whole contributions.

Each coefficient will be written as

\begin{equation}
{F_i}_{L,R}= \frac{\alpha^2}{2\,s_W^3\,m_W}
\sum_{d_i=d,\,s,\,b}V_{td_i}V^\dagger_{d_i c}
\left({G_i}_{L,R}^{tc+tc\gamma}+{G_i}_{L,R}^{tch}+\sum_{i=1}^3{G_i}_{L,R}^{{\rm
Box}_i}\right).
\end{equation}
The quark color factor is already included and we also introduced
explicitly the values of the quark charges. The nonzero coefficients
${G_i}_{L,R}^{tc+tc\gamma}$ are given by

\begin{eqnarray}
{G_4}_L^{tc+tc\gamma}&=&\frac{1}{\eta m_W^2}\Bigg[u m_t^2
\Big(3(C_0^{(1)} +C_1^{(1)}+ C_2^{(1)}-C_2^{(22)}
+C_{11}^{(1)}+2C_1^{(22)}+3C_{12}^{(1)} +
2C_{22}^{(1)})+2(C_2^{(13)}+C_{11}^{(21)})+  C_0^{(2)}\nonumber\\
&+& C_1^{(13)}+C_1^{(21)} + C_{12}^{(21)}
 \Big) - m_{d_i}^2\Big(C_1^{(21)} +
3(C_1^{(22)}+ C_{12}^{(1)}+ C_{22}^{(1)}+C_{12}^{(21)}) +
C_{11}^{(21)} \Big)- m_W^2\Big(3(C_0^{(1)}-C_1^{(1)}\nonumber\\ &-&
C_2^{(1)}+C_{12}^{(1)}+ C_{22}^{(1)}+2C_1^{(22)} ) +C_0^{(2)} +
C_1^{(21)} + 2( C_{11}^{(21)}  - C_{12}^{(21)} ) \Big)+
6C_{00}^{(1)} + 3C_{00}^{(21)}-B_0^{(1)}-B_0^{(2)} \Bigg],
\end{eqnarray}

\begin{eqnarray}
{G_1}_R^{tc+tc\gamma}&=&\frac{1}{2\eta m_W^2}\Bigg[
\frac{m_{d_i}^2}{ m_t^2} \Big(2(1 +B_0^{(1)}+ 2C_{00}^{(21)}
+3C_{00}^{(1)}) - 5B_0^{(2)} + B_0^{(7)}\Big)- u \Big( B_0^{(1)} -
B_0^{(2)}\Big) - u m_{d_i}^2C_2^{(13)}
\nonumber\\&+&\frac{\left({m_W^2}+m_{d_i}^2\right)m_{d_i}^2-
2m_W^4}{u m_t^4}\Big(B_0^{(2)} - B_0^{(7)}\Big) -\frac{2
m_W^2}{m_t^2} \Big(1 + 2(B_0^{(2)} - 3C_{00}^{(1)} - C_{00}^{(21)})
+ B_0^{(3)} - B_0^{(7)}\Big)
\nonumber\\
&+&u^2m_t^2 \Big(C_0^{(2)} + C_1^{(13)} + 2C_2^{(13)}\Big)-u m_W^2
\Big( C_0^{(2)} + 2(3(C_0^{(1)} + 2C_1^{(22)}+ 2C_2^{(22)}) +
C_2^{(13)} )\Big)\Bigg],
\end{eqnarray}

\noindent where $\xi=1-x$ and $\eta=1-u$. $B_0^{(i)}$, $C_0^{(i)}$,
and $D_0^{(i)}$ are Passarino-Veltman scalar functions
\cite{Passarino:1978jh}, whereas $C_{lm}^{(i)}$ and $D_{lm}^{(i)}$
stand for the coefficient functions of tensor integrals. We follow
the same nomenclature introduced in \cite{Mertig:1990an}. The
arguments of the Passarino-Veltman functions are represented by the
superscript $ (i)$ and are presented in Tables \ref{B0}, \ref{C0},
and \ref{D0}. Note that although the $C_0$ and $D_0$ scalar
functions are invariant under the permutation of their arguments,
this is not true  in general for the coefficient functions $C_{lm}$
and $D_{lm}$.

\begin{table}[!htb]
\caption{\label{B0}Arguments for the two-point Passarino-Veltman
scalar functions: $B_0^{(i)}=B_0(1,2,3)$. According to our notation
$(p_1-k)^2=m_t^2(1-x)$, $(p_1-p_2)^2=m_h^2 + m_t^2\,(x-u)$, and
$(p_1-q)^2=m_t^2\,u$.}
\begin{tabular}{cccc}
\hline $(i)$&1&2&3\\\hline
$(1)$&$0$&$m_{d_i}^2$&$m_{d_i}^2$\\
$(2)$&$0$&$m_{d_i}^2$&$m_W^2$\\
$(3)$&$0$&$m_W^2$&$m_W^2$\\
$(4)$&$m_h^2$&$m_{d_i}^2$&$m_{d_i}^2$\\
$(5)$&$m_h^2$&$m_W^2$&$m_W^2$\\
$(6)$&$m_t^2$&$m_{d_i}^2$&$m_W^2$\\
$(7)$&$(p_1-q)^2$&$m_{d_i}^2$&$m_W^2$\\
$(8)$&$(p_1-k)^2$&$m_{d_i}^2$&$m_W^2$\\
$(9)$&$(p_1-p_2)^2$&$m_{d_i}^2$&$m_{d_i}^2$\\
$(10)$&$(p_1-p_2)^2$&$m_W^2$&$m_W^2$\\
\hline
\end{tabular}
\end{table}

\begin{table}[!htb]
\caption{\label{C0}Arguments for the three-point Passarino-Veltman
coefficient functions: $C_{lm}^{(i)}=C_{lm}(1,2,3,4,5,6)$.}
\begin{tabular}{ccccccc}
\hline $(i)$&1&2&3&4&5&6\\\hline
$(1)$&$0$&$0$&$(p_1-q)^2$&$m_{d_i}^2$&$m_W^2$&$m_W^2$\\
$(2)$&$0$&$0$&$(p_1-q)^2$&$m_W^2$&$m_{d_i}^2$&$m_{d_i}^2$\\
$(3)$&$0$&$m_h^2$&$(p_1-k)^2$&$m_{d_i}^2$&$m_W^2$&$m_W^2$\\
$(4)$&$0$&$m_h^2$&$(p_1-k)^2$&$m_W^2$&$m_{d_i}^2$&$m_{d_i}^2$\\
$(5)$&$0$&$m_h^2$&$(p_1-p_2)^2$&$m_{d_i}^2$&$m_{d_i}^2$&$m_{d_i}^2$\\
$(6)$&$0$&$m_h^2$&$(p_1-p_2)^2$&$m_W^2$&$m_W^2$&$m_W^2$\\
$(7)$&$m_t^2$&$0$&$(p_1-k)^2$&$m_{d_i}^2$&$m_W^2$&$m_W^2$\\
$(8)$&$m_t^2$&$0$&$(p_1-k)^2$&$m_W^2$&$m_{d_i}^2$&$m_{d_i}^2$\\
$(9)$&$m_t^2$&$m_h^2$&$(p_1-q)^2$&$m_{d_i}^2$&$m_W^2$&$m_W^2$\\
$(10)$&$m_t^2$&$m_h^2$&$(p_1-q)^2$&$m_W^2$&$m_{d_i}^2$&$m_{d_i}^2$\\
$(11)$&$m_t^2$&$(p_1-p_2)^2$&$0$&$m_{d_i}^2$&$m_W^2$&$m_W^2$\\
$(12)$&$m_t^2$&$(p_1-p_2)^2$&$0$&$m_W^2$&$m_{d_i}^2$&$m_{d_i}^2$\\
$(13)$&$0$&$0$&$(p_1-q)^2$&$m_{d_i}^2$&$m_{d_i}^2$&$m_W^2$\\
$(14)$&$0$&$(p_1-k)^2$&$m_t^2$&$m_{d_i}^2$&$m_{d_i}^2$&$m_W^2$\\
$(15)$&$m_h^2$&$0$&$(p_1-k)^2$&$m_{d_i}^2$&$m_{d_i}^2$&$m_W^2$\\
$(16)$&$m_h^2$&$0$&$(p_1-p_2)^2$&$m_{d_i}^2$&$m_{d_i}^2$&$m_{d_i}^2$\\
$(17)$&$m_h^2$&$(p_1-q)^2$&$m_t^2$&$m_{d_i}^2$&$m_{d_i}^2$&$m_W^2$\\
$(18)$&$m_t^2$&$0$&$(p_1-p_2)^2$&$m_{d_i}^2$&$m_W^2$&$m_{d_i}^2$\\
$(19)$&$m_t^2$&$(p_1-q)^2$&$m_h^2$&$m_{d_i}^2$&$m_W^2$&$m_{d_i}^2$\\
$(20)$&$m_t^2$&$(p_1-k)^2$&$0$&$m_{d_i}^2$&$m_W^2$&$m_{d_i}^2$\\
$(21)$&$(p_1-q)^2$&$0$&$0$&$m_{d_i}^2$&$m_W^2$&$m_{d_i}^2$\\
$(22)$&$(p_1-q)^2$&$0$&$0$&$m_{d_i}^2$&$m_W^2$&$m_W^2$\\
$(23)$&$(p_1-q)^2$&$m_h^2$&$m_t^2$&$m_{d_i}^2$&$m_W^2$&$m_W^2$\\
$(24)$&$(p_1-k)^2$&$0$&$m_h^2$&$m_{d_i}^2$&$m_W^2$&$m_{d_i}^2$\\
$(25)$&$(p_1-k)^2$&$0$&$m_t^2$&$m_{d_i}^2$&$m_W^2$&$m_W^2$\\
$(26)$&$(p_1-k)^2$&$m_h^2$&$0$&$m_{d_i}^2$&$m_W^2$&$m_W^2$\\
$(27)$&$0$&$(p_1-q)^2$&$0$&$m_{d_i}^2$&$m_{d_i}^2$&$m_W^2$\\
$(28)$&$0$&$(p_1-k)^2$&$m_h^2$&$m_{d_i}^2$&$m_W^2$&$m_{d_i}^2$\\
\hline
\end{tabular}
\end{table}

\begin{table}[!htb]
\caption{\label{D0}Arguments for the four-point Passarino-Veltman
coefficient functions: $D_{lm}^{(i)}=D_{lm}(1,2,3,4,5,6,7,8,9,10)$.}
\begin{tabular}{ccccccccccc}
\hline $(i)$&1&2&3&4&5&6&7&8&9&10\\\hline
$(1)$&$0$&$(p_1-k)^2$&$m_h^2$&$(p_1-q)^2$&$m_t^2$&$0$&$m_W^2$&$m_W^2$&$m_{d_i}^2$&$m_{d_i}^2$\\
$(2)$&$m_h^2$&$(p_1-q)^2$&$0$&$(p_1-k)^2$&$m_t^2$&$0$&$m_W^2$&$m_W^2$&$m_{d_i}^2$&$m_{d_i}^2$\\
$(3)$&$m_t^2$&$0$&$0$&$m_h^2$&$(p_1-p_2)^2$&$(p_1-q)^2$&$m_{d_i}^2$&$m_W^2$&$m_{d_i}^2$&$m_{d_i}^2$\\
$(4)$&$m_t^2$&$0$&$m_h^2$&$0$&$(p_1-p_2)^2$&$(p_1-k)^2$&$m_{d_i}^2$&$m_W^2$&$m_{d_i}^2$&$m_{d_i}^2$\\
$(5)$&$(p_1-p_2)^2$&$0$&$(p_1-k)^2$&$0$&$m_t^2$&$m_h^2$&$m_W^2$&$m_W^2$&$m_{d_i}^2$&$m_W^2$\\
$(6)$&$(p_1-p_2)^2$&$m_t^2$&$(p_1-q)^2$&$0$&$0$&$m_h^2$&$m_W^2$&$m_W^2$&$m_{d_i}^2$&$m_W^2$\\
\hline
\end{tabular}
\end{table}

The remaining nonzero coefficients are:
\begin{eqnarray}
{G_2}_R^{tch}&=& \frac{1}{x}\Bigg[2\left(2\xi-\mu_h\right)
C_0^{(3)}+\frac{2m_{d_i}^2}{m_t^2m_W^2} \Big( 1 + B_0^{(1)} -
B_0^{(4)}
\Big)-2\left(\frac{m_{d_i}^2\mu_h}{m_W^2}-\frac{4m_W^2}{m_t^2}-\frac{m_h^2}{m_W^2}\left(\xi
-{\mu_h}\right)+2\mu_h\right)C_1^{(26)}
\nonumber\\
&-&2\frac{m_{d_i}^2}{m_W^2}\left(\frac{2m_{d_i}^2}{m_t^2}-\xi\right)\Big(C_0^{(4)}
+ C_1^{(15)} + C_2^{(15)}\Big)-\frac{2m_{d_i}^2}{m_t^2} \Big(
2C_0^{(3)} + 3C_0^{(4)}+4C_1^{(15)}+2C_1^{(26)} +
4C_2^{(15)}\Big)\nonumber\\
&-& \frac{2}{m_t^2} \Big(B_0^{(2)} + \ B_0^{(5)} -
2B_0^{(8)}\Big)+\frac{m_h^2}{m_t^2m_W^2} \Big(B_0^{(5)} -
2B_0^{(8)}\Big)\Bigg],
\end{eqnarray}

\begin{eqnarray}
{G_1}_R^{tch}&=&\frac{1}{u x m_W^2}\Bigg[\frac{m_h^2}{2 m_t^2}
\Big(x (B_0^{(5)} - 2 B_0^{(6)}) + u (2 B_0^{(8)} - B_0^{(5)}) \Big)
+ \frac{1}{2} x u B_0^{(7)}+ \frac{m_{d_i}^2}{2 m_t^2}\Big(2 (x -
u)(1 + B_0^{(1)} -
B_0^{(4)})\nonumber\\
&+& x (B_0^{(7)} - B_0^{(2)})\Big)-  \frac{m_{d_i}^2 m_h^2}{m_t^2}
\Big(u (C_0^{(4)} + C_1^{(15)} + C_2^{(15)} - C_1^{(26)}) - x
(C_0^{(10)} + C_1^{(17)} +
C_2^{(17)} - C_1^{(9)})\Big)\nonumber\\
&+& \frac{2 m_{d_i}^4}{m_t^2}\Big(u ( C_0^{(4)} + C_1^{(15)} +
C_2^{(15)}) - x ( C_0^{(10)} + C_1^{(17)} + C_2^{(17)})\Big)+
m_h^2\Big(x C_1^{(9)} - \mu_h (x
C_1^{(9)} - u C_1^{(26)}) \nonumber\\
&-& u ( C_1^{(26)} - x (C_0^{(9)} + C_1^{(9)} + C_1^{(26)} +
C_2^{(9)}))\Big)+ m_{d_i}^2\Big(x (C_0^{(10)} + C_1^{(17)} +
C_2^{(17)})
 - u (C_1^{(15)} +
C_2^{(15)}\nonumber\\
&+&  \xi C_0^{(4)} - x (C_0^{(10)} + C_1^{(17)}+ C_2^{(15)} +
C_2^{(17)} - C_1^{(15)}))\Big)+ 2m_W^2\Big(x C_0^{(9)} - u \xi
C_0^{(3)} - u x ( C_0^{(9)} + C_2^{(9)} ) \Big) \nonumber\\
&+& \frac{m_W^2}{2 m_t^2} \Big(x ( B_0^{(2)} - 3 B_0^{(7)} - 2
(B_0^{(5)} - 2 B_0^{(6)})) + 2 u (B_0^{(2)} + B_0^{(5)}- 2
B_0^{(8)})\Big)  + \frac{4
m_W^4}{m_t^2} \Big(u C_1^{(26)} - x C_1^{(9)}\Big)\nonumber\\
&-& \frac{m_h^2m_W^2}{m_t^2} \Big(x ( C_0^{(9)} - 2 (C_1^{(9)} -2
C_2^{(9)})) - u ( C_0^{(3)} + 2 (2C_2^{(26)} - C_1^{(26)}))\Big)+
\frac{m_{d_i}^2m_W^2}{m_t^2}\Big(u (3 C_0^{(4)} \nonumber\\
&+& 2 (C_0^{(3)} + C_1^{(26)} + 2 (C_1^{(15)} + C_2^{(15)}))) - x (3
C_0^{(10)} + 2 (C_0^{(9)} - C_1^{(9)}+ 2 (C_1^{(17)} +
C_2^{(17)})))\Big)\Bigg],
\end{eqnarray}


\begin{eqnarray}
{G_3}_L^{\rm Box_1}&=&\frac{m_{d_i}^2}{m_W^2}\Bigg[\Big( C_1^{(14)}
+ C_{12}^{(14)} - C_0^{(4)} - C_0^{(8)} - C_1^{(16)} - C_1^{(20)} -
C_1^{(24)} - C_2^{(5)} -
C_2^{(16)} - C_2^{(20)} - C_2^{(24)}\Big)\nonumber\\
&+& m_h^2 \Big(D_0^{(3)} + D_0^{(4)} + 2(D_1^{(3)}  + D_{12}^{(3)} +
D_{12}^{(4)}+D_2^{(3)} + D_2^{(4)} - D_{13}^{(3)} + D_3^{(3)} +
D_{23}^{(3)}) +
D_{22}^{(3)}+D_{22}^{(4)}+ D_{33}^{(3)} \Big) \nonumber\\
&-&2 m_{d_i}^2 \Big(D_0^{(3)} + D_0^{(4)} + D_1^{(3)} + D_1^{(4)}+
D_2^{(3)} + D_2^{(4)} + D_{13}^{(3)} + D_{33}^{(3)} - D_{13}^{(4)}-
D_{23}^{(3)} - D_{23}^{(4)} + 2
D_3^{(3)}\Big)\nonumber\\&-&m_t^2\Big(D_0^{(3)} + D_{13}^{(4)} +
D_{23}^{(4)} + D_{33}^{(3)} + 2 (D_3^{(3)} + D_{11}^{(3)} +
D_{11}^{(4)}+ D_{22}^{(3)} + D_{22}^{(4)} + 2 (D_{12}^{(3)}+
D_{12}^{(4)})) + 3 (D_1^{(3)} + D_1^{(4)}\nonumber\\ &+& D_2^{(3)} +
D_2^{(4)} + D_{13}^{(3)} + D_{23}^{(3)})+ (u + \xi) D_0^{(4)}- x
(D_1^{(4)} - D_2^{(4)})+ u (D_1^{(3)} + D_2^{(4)} + D_3^{(4)} +
D_{11}^{(4)} + D_{12}^{(3)} +D_{12}^{(4)} \nonumber\\&+&
D_{13}^{(4)} +D_{23}^{(4)} -D_3^{(3)} - D_{11}^{(3)} - D_{23}^{(3)}-
D_{33}^{(3)} + 2 D_1^{(4)})\Big) + m_W^2 \Big(D_0^{(3)} + D_0^{(4)}
-
D_3^{(3)} + 3 (D_1^{(3)} + D_1^{(4)}+D_2^{(3)} \nonumber\\
&+&  D_2^{(4)}) - 4 (D_{13}^{(3)}+D_{23}^{(3)} + D_{33}^{(3)} -
D_{13}^{(4)} - D_{23}^{(4)}) \Big)\Bigg],
\end{eqnarray}

\begin{eqnarray}
{G_4}_L^{\rm Box_1}&=& \frac{m_{d_i}^2}{m_W^2}\Bigg[ C_1^{(16)} +
C_2^{(5)} + C_2^{(16)} + C_2^{(24)} - C_{12}^{(27)} + 2 (C_0^{(5)} -
D_{00}^{(3)} - D_{00}^{(4)})- m_h^2
\Big(D_3^{(3)} + D_{22}^{(3)} \nonumber\\
&+& D_{22}^{(4)} + D_{33}^{(3)} + 3 D_{23}^{(3)}\Big) + 2
m_{d_i}^2\Big(D_3^{(3)} + D_{23}^{(3)} + D_{33}^{(3)}+
D_{23}^{(4)}\Big)+ m_t^2\Big(D_3^{(3)} + D_{23}^{(4)} +
D_{33}^{(3)}\nonumber\\
&+& 3 D_{23}^{(3)} + 2 (D_2^{(3)} + D_{12}^{(4)} + D_{13}^{(3)} +
D_{22}^{(4)} - D_{22}^{(3)})+(2 + u - x)
D_2^{(4)} + (2 + u) D_{12}^{(3)} \nonumber\\
&+& u ( D_{12}^{(4)} + D_{13}^{(3)} + D_{23}^{(4)} - D_{23}^{(3)} -
D_{33}^{(3)}) - \mu_h (D_2^{(3)}+ D_2^{(4)} + D_{12}^{(3)} +
D_{12}^{(4)}  -
D_{13}^{(3)} - D_{23}^{(3)})\Big)\nonumber\\
&+&  m_W^2 \Big(D_2^{(3)} + D_2^{(4)} + 5 D_3^{(3)}+ 2 (D_0^{(3)} +
D_0^{(4)} + 2 (D_{23}^{(3)} + D_{33}^{(3)} - D_{23}^{(4)}))
\Big)\Bigg],
\end{eqnarray}

\begin{eqnarray}
{G_2}_R^{\rm Box_1}&=& -\frac{m_{d_i}^2}{m_W^2}\Bigg[ C_0^{(4)} +
C_0^{(5)} + C_1^{(16)} + C_1^{(20)} + C_1^{(24)} + C_2^{(5)} +
C_2^{(16)} + C_2^{(24)} + C_{22}^{(14)} - 2
C_2^{(14)}-2 m_{d_i}^2 \Big(D_0^{(3)}+ D_1^{(4)}\nonumber\\
&+& D_2^{(4)} + D_{11}^{(3)} + D_{11}^{(4)} +D_{22}^{(3)} +
D_{22}^{(4)} + D_{33}^{(3)}+  2 (D_1^{(3)} + D_2^{(3)}+ D_3^{(3)} +
D_{12}^{(4)}+ D_{13}^{(3)}+ D_{23}^{(3)} -
D_{12}^{(3)})\Big)\nonumber\\&-& m_W^2 \Big(3 (D_0^{(3)} + D_1^{(4)}
+ D_2^{(4)}) + 7 (D_1^{(3)}+ D_2^{(3)} + D_3^{(3)})+ 4 (D_{11}^{(3)}
+ D_{22}^{(3)} + D_{22}^{(4)} -
D_{11}^{(4)} - D_{33}^{(3)} + 2 (D_{12}^{(3)}\nonumber\\
&+& D_{12}^{(4)} + D_{13}^{(3)} + D_{23}^{(3)}))\Big)+
m_t^2\Big(D_0^{(3)} + D_1^{(4)} + D_{11}^{(3)} + D_{11}^{(4)} +
D_{22}^{(3)} + D_{22}^{(4)} + D_{33}^{(3)} +
2 (D_1^{(3)} + D_3^{(3)} + D_{12}^{(3)}\nonumber\\
&+& D_{12}^{(4)} + D_{13}^{(3)} + D_{23}^{(3)}) + \eta
 D_2^{(4)} - (u - 2) D_2^{(3)} - u (D_3^{(3)} +
D_{12}^{(3)} + D_{12}^{(4)} + D_{13}^{(3)} + D_{22}^{(3)}+
D_{22}^{(4)} + D_{33}^{(3)}\nonumber\\ &-& 2 D_{23}^{(3)}) \Big)
-4m_h^2 D_2^{(3)}\Bigg],
\end{eqnarray}

\begin{eqnarray}
{G_1}_R^{\rm Box_1}&=& \frac{m_{d_i}^2}{2 m_W^2}\Bigg[ C_0^{(8)} +
C_0^{(10)} + C_1^{(20)} + C_2^{(17)} + 2 (C_0^{(5)} + C_1^{(17)}
- D_{00}^{(3)} - D_{00}^{(4)}) -2 m_h^2D_0^{(4)}\nonumber\\
&+& (2m_{d_i}^2-m_t^2) \Big(D_0^{(3)} + D_0^{(4)} + D_1^{(3)} +
D_1^{(4)} + D_2^{(3)}+ D_2^{(4)} + D_3^{(3)}\Big)+ m_t^2\Big(u
D_3^{(3)}-x D_3^{(4)}\Big)\nonumber\\ &-& m_W^2\Big(D_0^{(3)} +
D_0^{(4)}+ 4 ( D_1^{(3)} + D_1^{(4)} + D_2^{(3)} + D_2^{(4)} +
D_3^{(3)})\Big)\Bigg],
\end{eqnarray}


\begin{eqnarray}
{G_3}_L^{\rm Box_2}&=& -\frac{m_{d_i}^2}{2 m_W^2}\Bigg[ 2
(C_1^{(14)} + C_2^{(14)} + C_{12}^{(14)} - C_1^{(20)} - C_2^{(20)} -
3 (C_1^{(19)} +
C_2^{(25)} + C_{12}^{(25)} + D_{00}^{(1)}-2 C_1^{(25)})) \nonumber\\
&+& 3 ( C_2^{(19)} + C_2^{(28)} - C_0^{(10)} + 2 (C_0^{(7)} +
C_{11}^{(25)})) + m_t^2 \Big(3 (( x -2 (u + 1))
D_2^{(1)}+ (x - 2) ( D_{12}^{(1)} + D_{23}^{(1)}\nonumber\\
&+& 2 D_{22}^{(1)}) + u (D_3^{(1)} + D_{13}^{(1)} + D_{33}^{(1)}- 2
(D_{12}^{(1)} + D_{22}^{(1)} + D_{23}^{(1)} )))-\mu_h (D_1^{(2)} +
D_3^{(2)} + D_{11}^{(2)} \nonumber\\&+& D_{23}^{(2)}+D_{33}^{(2)} -
D_{12}^{(2)} + 2 D_{13}^{(2)}- 3 (D_{22}^{(1)}- D_1^{(1)}))\Big) +
12 m_{d_i}^2 \Big(D_2^{(1)} + D_{12}^{(1)} +
D_{23}^{(1)}\Big)\Bigg]+C_2^{(14)}- C_1^{(9)}\nonumber\\
&-& 2 m_t^2 \Big(D_{11}^{(2)} + D_{33}^{(2)}+ 2 (D_3^{(2)} +
D_{13}^{(2)} + D_{22}^{(2)}) + 3 (D_2^{(2)} + D_{12}^{(2)}
+ D_{23}^{(2)}) + \xi (D_0^{(2)} + 2 D_1^{(2)})\nonumber\\
&+& u (D_2^{(2)} + D_{12}^{(2)}+ D_{22}^{(2)} + D_{23}^{(2)}) - x
(D_{11}^{(2)} + D_{33}^{(2)} - D_{22}^{(2)} + 2 (D_2^{(2)} +
D_3^{(2)} + D_{12}^{(2)} +D_{13}^{(2)}+ D_{23}^{(2)}))\nonumber\\
&-& \mu_h ( D_0^{(2)} + D_{11}^{(2)} + D_{22}^{(2)} + D_{33}^{(2)} +
2 (D_1^{(2)} + D_2^{(2)} + D_{12}^{(2)} +
D_{13}^{(2)} + D_{23}^{(2)} - D_3^{(2)}))\Big)\nonumber\\
&-& 2 m_W^2 \Big(2 (D_{11}^{(2)} + D_{12}^{(2)} + D_{33}^{(2)} -
D_{23}^{(2)} + 2 D_{13}^{(2)}) + 3 (D_0^{(2)} + D_2^{(2)}) + 5
(D_1^{(2)} + D_3^{(2)})\Big)\nonumber\\&+&2m_{d_i}^2 \Big(2
(D_2^{(2)} - D_{11}^{(2)} - D_{12}^{(2)} - D_{23}^{(2)}-
D_{33}^{(2)} - 2 (D_{13}^{(2)} - 3 (D_{12}^{(1)} + D_{23}^{(1)})))-
3 (2 D_0^{(1)} + 3 D_2^{(1)})\Big) ,
\end{eqnarray}

\begin{eqnarray}
{G_4}_L^{\rm Box_2}&=& \frac{m_{d_i}^2}{2 m_W^2}\Bigg[3 (C_0^{(4)} +
C_2^{(28)} - C_2^{(19)} + 2 (C_2^{(1)}+C_{12}^{(1)}+C_{22}^{(1)} - 3
D_{00}^{(1)}))- 2 (C_2^{(13)} + C_{12}^{(13)} +
C_{22}^{(13)})\nonumber\\
&+&12 m_{d_i}^2 \Big(D_3^{(1)} + D_{13}^{(1)} + D_{33}^{(1)}\Big)+
2m_W^2 \Big(2 (D_{11}^{(2)} + D_{13}^{(2)} + 6 (D_{13}^{(1)} +
D_{33}^{(1)})) + 3 (2 D_0^{(1)} + 5 D_3^{(1)})\Big)\nonumber\\&-&
m_t^2 \Big(3 (2 ( 1 + u )
 D_3^{(1)} + ( 2 + 3 u )  (D_{13}^{(1)} +
D_{33}^{(1)}) + x D_{12}^{(1)} +(2(2 + u) - x)
D_{23}^{(1)})- 2\mu_h (D_1^{(2)} + D_{11}^{(2)}\nonumber\\
&+& D_{13}^{(2)} + 3 (D_{23}^{(1)} -2 D_{13}^{(1)})) \Big)\Bigg] + 2
m_t^2 \Big(D_2^{(2)} + D_3^{(2)} + D_{11}^{(2)}+ D_{13}^{(2)} + 2
D_{12}^{(2)} + \xi (D_0^{(2)} + 2 D_1^{(2)})\nonumber\\
&+& u (D_2^{(2)} + D_{12}^{(2)})- x (D_2^{(2)} + D_3^{(2)} +
D_{11}^{(2)} + D_{12}^{(2)} + D_{13}^{(2)} )+ {\mu_h} (D_0^{(2)}
+ D_2^{(2)} + D_3^{(2)} + D_{11}^{(2)}+ D_{12}^{(2)}\nonumber\\
&+& D_{13}^{(2)}+ 2 D_1^{(2)})\Big) + 2 m_W^2 \Big(D_0^{(2)} +
D_1^{(2)} + 2 (D_{11}^{(2)} + D_{13}^{(2)} )\Big)+  2
\Big(C_0^{(8)}-C_0^{(9)} - C_2^{(9)}-2 D_{00}^{(2)}\Big),
\end{eqnarray}

\begin{eqnarray}
{G_2}_R^{\rm Box_2}&=& \frac{m_{d_i}^2}{2
m_W^2}\Bigg[6\left(2m_{d_i}^2 - m_h^2\right) D_{22}^{(1)}+ 2
(C_1^{(20)} - C_{22}^{(14)} + 3 C_{22}^{(25)} - 2 ( C_2^{(14)}
- 3 C_1^{(25)})) + 3 (C_0^{(10)} + C_1^{(19)}\nonumber\\
&+& C_2^{(19)} - C_2^{(28)}+ 2 (C_0^{(7)} + C_{11}^{(25)} - 2
C_2^{(25)} + 3 C_{12}^{(25)})) - 2m_{d_i}^2 m_t^2 \Big(\mu_h
(C_1^{(9)} - C_1^{(20)})+ \xi (C_1^{(20)}\nonumber\\ &-& C_2^{(14)})
- u (C_1^{(14)} - C_1^{(20)} - C_2^{(14)} - C_2^{(20)})\Big)- m_t^2
\Big(3 ( D_{22}^{(1)} + (2 - x) D_{22}^{(1)} + u D_{23}^{(1)})- 2
\mu_h ( D_0^{(2)} \nonumber\\&+& D_{11}^{(2)} + D_{22}^{(2)} -
D_{33}^{(2)} + 2 (D_1^{(2)}+ D_2^{(2)} + D_{12}^{(2)} + D_{13}^{(2)}
+ D_{23}^{(2)} - D_3^{(2)}))\Big)+2 m_W^2 \Big(3
D_2^{(1)}\nonumber\\&+& 2 (D_1^{(2)} + D_2^{(2)}+D_3^{(2)} +
D_{11}^{(2)} + D_{22}^{(2)} + D_{33}^{(2)} + 2 (D_{12}^{(2)} +
D_{13}^{(2)} + D_{23}^{(2)}- 3 D_{22}^{(1)}))\Big)\Bigg]-
2(C_0^{(3)} \nonumber\\ &-& C_1^{(9)}+C_1^{(26)} + C_2^{(14)})-
\frac{m_h^2 m_t^2}{m_W^2} \Big(D_{11}^{(2)} + D_{22}^{(2)} +
D_{33}^{(2)} + 2 (D_2^{(2)} + D_3^{(2)}+
D_{12}^{(2)} + D_{13}^{(2)} + D_{23}^{(2)})\nonumber\\
&+& \xi (D_0^{(2)} + 2 D_1^{(2)})+ u (D_2^{(2)} + D_{12}^{(2)} +
D_{22}^{(2)} + D_{23}^{(2)}) - x ( D_{11}^{(2)}+ D_{33}^{(2)} -
D_{22}^{(2)} + 2 (D_2^{(2)} + D_3^{(2)} \nonumber\\&+&D_{12}^{(2)} +
D_{13}^{(2)} + D_{23}^{(2)})) - \mu_h (D_0^{(2)} + D_{11}^{(2)} +
D_{22}^{(2)}+ D_{33}^{(2)} + 2 (D_1^{(2)} + D_2^{(2)} + D_{12}^{(2)}
+ D_{13}^{(2)}
 \nonumber\\
&+&D_{23}^{(2)} - D_3^{(2)}))\Big)+  m_t^2 \Big(2 (D_1^{(2)} +
D_3^{(2)} + \xi D_0^{(2)} - (1 + u) D_2^{(2)} + u (D_{11}^{(2)} +
D_{13}^{(2)} + D_{22}^{(2)} + D_{23}^{(2)}\nonumber\\&+& 2
D_{12}^{(2)}) - x (D_{11}^{(2)}+ D_{33}^{(2)} - D_{22}^{(2)} + 2
(D_1^{(2)} + D_2^{(2)} + D_3^{(2)} + D_{12}^{(2)} + D_{13}^{(2)} +
D_{23}^{(2)})))\nonumber\\&-& {\mu_h} ( 3 D_0^{(2)} + 2
(D_{11}^{(2)} + D_{22}^{(2)}+ D_{33}^{(2)} + 2 (D_{12}^{(2)} +
D_{23}^{(2)} - D_{13}^{(2)})) + 5 (D_1^{(2)} + D_2^{(2)} +
D_3^{(2)}))\Big)\nonumber\\ &+& 4 m_W^2 \Big(D_0^{(2)} +
D_{11}^{(2)}+ D_{33}^{(2)} - D_{22}^{(2)} + 2 (D_1^{(2)} + D_2^{(2)}
+ D_3^{(2)} + D_{12}^{(2)} + D_{13}^{(2)} + D_{23}^{(2)})\Big),
\end{eqnarray}

\begin{eqnarray}
{G_1}_R^{\rm Box_2}&=&\frac{1}{4m_W^2}\Bigg[  m_{d_i}^2 \Big(3
(C_0^{(4)}-C_0^{(10)}+ 2 (C_2^{(28)} - C_1^{(19)} - C_2^{(19)} - 4
D_{00}^{(1)}))-2
(C_0^{(8)} + C_1^{(20)}) \Big)+B_0^{(7)}-2 B_0^{(1)}  \nonumber\\
&+& 2m_t^2 \Big({\mu_h} (C_0^{(8)} + C_2^{(14)}
-C_0^{(9)}-C_1^{(20)} - C_2^{(9)}) - u (C_0^{(2)} + C_1^{(13)} +
C_1^{(20)} + C_2^{(13)} + C_2^{(20)} -C_1^{(14)} -
C_2^{(14)})\nonumber\\&+&\xi
 (C_1^{(20)}-C_2^{(14)} )\Big)
+ m_{d_i}^2 m_t^2\Big(\mu_h (3 D_0^{(1)} - 2 (D_0^{(2)}+ D_1^{(2)} +
D_2^{(2)} + D_3^{(2)} - 3 D_2^{(1)}))- 3 ( x ( D_2^{(1)} - 2 (
D_{12}^{(1)}\nonumber\\ &+& D_{23}^{(1)}))+ u ( D_3^{(1)} + 2 (
D_{13}^{(1)} + D_{33}^{(1)})))\Big) + 2m_h^2 m_t^2 \Big(\xi
(D_0^{(2)} + D_1^{(2)}+ D_3^{(2)})+ (u + \xi) D_2^{(2)} - \mu_h (
D_0^{(2)} + D_2^{(2)}\nonumber\\&+&D_3^{(2)} - D_1^{(2)})\Big) +
8m_W^4 \Big(D_0^{(2)} + D_1^{(2)}+ D_2^{(2)} + D_3^{(2)}\Big) +
2m_W^2 \Big(2 (C_0^{(9)} +
C_1^{(9)} - C_0^{(8)} - 2 D_{00}^{(2)})+C_0^{(2)} \Big)\nonumber\\
&-&2 m_{d_i}^2 m_W^2 \Big(D_1^{(2)} + D_2^{(2)} + D_3^{(2)} - 6
(D_0^{(1)} + 2 D_2^{(1)})\Big)- 2m_W^2m_t^2 \Big(2 \xi D_0^{(2)} - 2
(x - u) D_2^{(2)}\nonumber\\ &-& \mu_h (3 D_0^{(2)} + 2 ( D_1^{(2)}
+ D_2^{(2)} - D_3^{(2)}))\Big)+ \frac{m_{d_i}^2-m_W^2}{u m_t^2}
\Big(B_0^{(2)} - B_0^{(7)}\Big) - 3 m_{d_i}^2 m_h^2 \Big( D_0^{(1)}
+ 2 D_2^{(1)}\Big) \Bigg].
\end{eqnarray}


\begin{eqnarray}
{G_3}_L^{\rm Box_3}&=& 6\Bigg[ \frac{m_{d_i}^2}{2m_W^2} \Big(
C_1^{(25)} + C_{11}^{(25)} + C_{12}^{(25)}\Big)+ m_h^2
\Big(D_{11}^{(5)}+ D_{11}^{(6)}+D_0^{(5)} + 2(D_2^{(5)}+D_{12}^{(5)}
+D_1^{(5)})\Big) \nonumber\\
&+& \Big(C_1^{(25)} + C_1^{(26)} + C_2^{(23)} + C_2^{(25)} -
C_0^{(6)} - C_1^{(9)} + 2 C_{12}^{(6)}\Big) + m_{d_i}^2
\Big(D_1^{(5)} +
D_2^{(5)} + D_{13}^{(5)}+D_{13}^{(6)}\nonumber\\
&+&D_{23}^{(5)}-  D_0^{(6)} - D_1^{(6)}\Big) - m_t^2 \Big(D_1^{(6)}
+ D_{13}^{(5)} + D_{13}^{(6)} + D_{23}^{(5)} +2 (D_{11}^{(5)} +
D_{22}^{(5)}
-D_{11}^{(6)} + 2 D_{12}^{(5)}) \nonumber\\
&+& 3 (D_1^{(5)} + D_2^{(5)}) + (u + \xi) D_0^{(5)} - x (D_1^{(5)} +
D_2^{(5)}) - {\mu_h} D_{22}^{(5)}- u (D_1^{(5)} + D_3^{(5)} +
D_{12}^{(5)} +
D_{13}^{(5)}\nonumber\\
&+&D_{13}^{(6)} + D_{22}^{(5)}+D_{23}^{(5)} - D_{12}^{(6)} + 2
D_2^{(5)})\Big)+ m_W^2 \Big(2 (D_{13}^{(5)} + D_{13}^{(6)} +
D_{23}^{(5)}) - 3 (D_0^{(5)} + D_1^{(5)} \nonumber\\&+& D_2^{(5)}-
D_1^{(6)})\Big)+\frac{m_h^2}{m_W^2}
 C_{12}^{(6)} +
\frac{m_{d_i}^2 m_h^2}{m_W^2} \Big( D_{13}^{(5)} + D_{13}^{(6)} +
D_{23}^{(5)}\Big)\Bigg],
\end{eqnarray}

\begin{eqnarray}
{G_4}_L^{\rm Box_3}&=&-6\Bigg[ \frac{m_h^2}{m_W^2}
 C_{12}^{(6)}- \Big(C_0^{(3)} + C_0^{(6)} +
C_0^{(7)} + C_1^{(23)} + C_2^{(26)} - C_2^{(9)} - 2
(C_{12}^{(6)} + D_{00}^{(5)} - D_{00}^{(6)})\Big)\nonumber\\
&+& m_h^2 \Big(D_0^{(5)} + D_{11}^{(5)} + D_{11}^{(6)}+
D_{12}^{(6)}+ 2( D_1^{(5)}-D_{22}^{(6)})\Big) + \frac{ m_{d_i}^2
m_h^2}{2m_W^2}\Big( D_{13}^{(5)} + D_{13}^{(6)} +
D_{23}^{(6)}\Big)\nonumber\\&-& m_{d_i}^2 \Big(D_0^{(6)} + D_1^{(6)}
+ D_2^{(6)} - D_1^{(5)} - D_{13}^{(5)} - D_{13}^{(6)} -
D_{23}^{(6)}\Big) - m_t^2 \Big(D_1^{(6)}
+D_2^{(5)} + D_{13}^{(5)}+ D_{13}^{(6)}\nonumber\\
&+&  D_{23}^{(6)} + 2 (D_{11}^{(5)} + D_{11}^{(6)} + D_{12}^{(6)} -
D_{12}^{(5)}) + (u + \xi) D_0^{(5)} + (3 + u - x) D_1^{(5)} -\mu_h
(D_2^{(5)}
- D_{12}^{(5)})\nonumber\\
&+& u (D_2^{(5)} + D_{12}^{(5)} + D_{13}^{(5)}+ D_{13}^{(6)} +
D_{23}^{(6)} - D_3^{(5)} - D_{12}^{(6)} - D_{22}^{(6)})\Big)- m_W^2
\Big(D_0^{(5)} + D_1^{(6)} + D_2^{(6)}\nonumber\\ &-& D_1^{(5)} + 2
(D_0^{(6)}- D_{13}^{(5)} - D_{13}^{(6)} -
D_{23}^{(6)})\Big)+\frac{m_{d_i}^2}{2m_W^2} \Big( C_2^{(1)} +
C_{12}^{(1)} + C_{22}^{(1)}\Big) \Bigg],
\end{eqnarray}

\begin{eqnarray}
{G_2}_R^{\rm Box_3}&=& \frac{3}{m_W^2}\Bigg[ m_h^2\Big(C_0^{(3)}+
C_0^{(7)} + C_1^{(25)} +C_1^{(26)} +C_2^{(6)} + C_2^{(23)} +
C_2^{(25)} - C_1^{(9)} + 2 (C_2^{(11)} -
C_{22}^{(6)})\Big)\nonumber\\&-& \frac{1}{2}\Big(B_0^{(8)} + 2
(B_0^{(6)} - B_0^{(10)} - 2 C_{00}^{(25)})\Big)+ 4 m_W^4 \Big(
D_1^{(5)} + D_2^{(5)} + D_{11}^{(5)} +
D_{11}^{(6)} + D_{22}^{(5)}+ 2 D_{12}^{(5)}\Big) \nonumber\\
&+& m_{d_i}^2\Big(C_0^{(11)} - C_1^{(25)} - C_2^{(25)} -
C_{11}^{(25)} - C_{22}^{(25)}- 2 C_{12}^{(25)}\Big) + m_{d_i}^2
m_h^2\Big(D_1^{(5)} + D_2^{(5)} +
D_{11}^{(5)}+D_{11}^{(6)}\nonumber\\ &+& D_{22}^{(5)} + 2
D_{12}^{(5)}\Big) +m_t^2\Big(C_0^{(7)}+ 3 (C_1^{(25)} - C_2^{(25)})
+ (u + \xi) C_1^{(11)} - \mu_h (C_0^{(11)} + C_1^{(11)} + 2
C_2^{(11)})\nonumber\\
&-& x (C_1^{(25)} + C_{11}^{(25)} + C_{12}^{(25)})+ 2 (C_{11}^{(25)}
- C_{22}^{(25)} + 2 C_{12}^{(25)})\Big)+6 m_h^4\Big(D_1^{(5)}+
D_2^{(5)} + D_{11}^{(5)} + D_{11}^{(6)} + 2D_{12}^{(5)} \Big)
\nonumber\\ &-& m_h^2 m_t^2\Big(D_{11}^{(5)} + D_{11}^{(6)} -
D_{22}^{(5)} + 2 D_{12}^{(5)} + (u + \xi) (D_1^{(5)}+ D_2^{(5)}) -
\mu_h D_{22}^{(5)}  + u (D_{12}^{(5)} + D_{13}^{(6)} \nonumber\\&+&
D_{22}^{(5)} + D_{23}^{(5)} - D_{12}^{(6)} -D_{13}^{(5)})\Big)+
m_W^2\Big(C_0^{(11)} + 2 (C_2^{(6)} + 2 C_{22}^{(6)}) - 3 (C_0^{(7)}
+ C_1^{(25)} + C_2^{(25)})\Big)\nonumber\\ &+& m_h^2
m_w^2\Big(D_1^{(6)}- 2 ( 2 D_{12}^{(5)} + 3 D_{11}^{(5)} +
D_{11}^{(6)} - D_{22}^{(5)}) - 3 (D_1^{(5)} + D_2^{(5)})\Big) + 2
m_{d_i}^2 m_W^2 \Big(D_1^{(6)} +
D_{11}^{(5)}+ D_{11}^{(6)} \nonumber\\
&+& D_{22}^{(5)} +2 D_{12}^{(5)}\Big) + 2 m_t^2 m_W^2 \Big(D_1^{(5)}
+ D_2^{(5)} - D_1^{(6)} - x (D_1^{(5)} + D_2^{(5)})  + u (D_0^{(6)}
+ D_1^{(5)}+ D_2^{(5)}+D_{11}^{(5)} + D_{11}^{(6)}\nonumber\\
&+& D_{13}^{(5)} + D_{13}^{(6)}+D_{22}^{(5)} + D_{23}^{(5)} -
D_2^{(6)} + 2 (D_1^{(6)} +
D_{12}^{(5)}))\Big)+\frac{m_{d_i}^2-m_W^2}{2 \xi
 m_t^2}\Big(B_0^{(8)}-B_0^{(2)}\Big)\Bigg],
\end{eqnarray}

\begin{eqnarray}
{G_1}_R^{\rm Box_3}&=&3 \Bigg[ \frac{m_{d_i}^2-m_W^2}{4 u m_t^2
m_W^2} \Big(B_0^{(7)} - B_0^{(2)}\Big)+ \frac{1}{4 m_W^2}
 B_0^{(7)} + \frac{m_h^2}{2 m_W^2}  C_1^{(23)}  - \Big(
C_2^{(23)} +
2 ( D_{00}^{(5)} + D_{00}^{(6)} )\Big)\nonumber\\
&+& m_t^2 ( u - x )  D_3^{(6)}+ 4 m_W^2 \Big(D_0^{(5)} + D_1^{(5)} -
D_1^{(6)} + D_2^{(5)}\Big)\Bigg],
\end{eqnarray}



\begin{thebibliography}{99}

\bibitem{Chakraborty:2003iw}
D.~Chakraborty, J.~Konigsberg, and D.~L.~Rainwater,
Ann.\ Rev.\ Nucl.\ Part.\ Sci.\  {\bf 53}, 301 (2003);
M.~Beneke {\it et al.},
arXiv:hep-ph/0003033.


\bibitem{topcolor} S.~Weinberg,
Phys.\ Rev.\ D {\bf 13}, 974 (1976); L.~Susskind,
Phys.\ Rev.\ D {\bf 20}, 2619 (1979); C.~T.~Hill,
Phys.\ Lett.\ B {\bf 345}, 483 (1995); K.~D.~Lane,
Phys.\ Lett.\ B {\bf 433}, 96 (1998).


\bibitem{Mahlon:1994us}
G.~Mahlon and S.~J.~Parke,
Phys.\ Lett.\ B {\bf 347}, 394 (1995).

\bibitem{Decker:1992wz}
R.~Decker, M.~Nowakowski, and A.~Pilaftsis,
Z.\ Phys.\ C {\bf 57}, 339 (1993).

\bibitem{Altarelli:2000nt}
G.~Altarelli, L.~Conti, and V.~Lubicz,
Phys.\ Lett.\ B {\bf 502}, 125 (2001).


\bibitem{Jenkins:1996zd}
E.~Jenkins,
Phys.\ Rev.\ D {\bf 56}, 458 (1997).


\bibitem{Han:1995pk}
T.~Han, R.~D.~Peccei, and X.~Zhang,
Nucl.\ Phys.\ B {\bf 454}, 527 (1995).

\bibitem{Eilam:1990zc}
G.~Eilam, J.~L.~Hewett, and A.~Soni,
Phys.\ Rev.\ D {\bf 44}, 1473 (1991); {\it ibid.} D {\bf 59}, 039901
(E) (1999).

\bibitem{Mele:1999zk}
B.~Mele, S.~Petrarca, and A. Soddu,
Phys. Lett. {\bf B435}, 401 (1999).

\bibitem{Hou:1991un}
W.~S.~Hou,
Phys.\ Lett.\ B {\bf 296}, 179 (1992); see also E.~O.~Iltan,
Phys.\ Rev.\ D {\bf 65}, 075017 (2002).

\bibitem{Yang:1993rb}
J.~M.~Yang and C.~S.~Li,
Phys.\ Rev.\ D {\bf 49}, 3412 (1994); {\it ibid.} D {\bf 51}, 3974
(E) (1995).

\bibitem{Eilam:2001dh}
G.~Eilam, A.~Gemintern, T.~Han, J.~M.~Yang, and X.~Zhang,
Phys.\ Lett.\ B {\bf 510}, 227 (2001).

\bibitem{Guasch:1999jp}
J.~Guasch and J.~Sola,
Nucl.\ Phys.\ B {\bf 562}, 3 (1999).

\bibitem{Bejar:2000ub}
S.~Bejar, J.~Guasch and J.~Sola,
Nucl.\ Phys.\ B {\bf 600}, 21 (2001).


\bibitem{Diaz1} J. L. D\'{\i}az-Cruz, R. Martinez, M.A. P\'erez, and A.
Rosado, Phys. Rev. D {\bf{41}}, 891 (1990).


\bibitem{deDivitiis:1997sh}
G.~M.~de Divitiis, R.~Petronzio, and L.~Silvestrini,
Nucl.\ Phys.\ B {\bf 504}, 45 (1997).

\bibitem{Couture:1994rr}
G.~Couture, C.~Hamzaoui, and H.~Konig,
Phys.\ Rev.\ D {\bf 52}, 1713 (1995).

\bibitem{Li:1993mg}
C.~S.~Li, R.~J.~Oakes, and J.~M.~Yang,
Phys.\ Rev.\ D {\bf 49}, 293 (1994); {\it ibid.} D {\bf 56}, 3156
(E) (1997).


\bibitem{Yang:1997dk}
J.~M.~Yang, B.~L.~Young, and X.~Zhang,
Phys.\ Rev.\ D {\bf 58}, 055001 (1998).

\bibitem{Lopez:1997xv}
J.~L.~Lopez, D.~V.~Nanopoulos, and R.~Rangarajan,
Phys.\ Rev.\ D {\bf 56}, 3100 (1997).

\bibitem{Yue:2001qr}
C.~x.~Yue, G.~r.~Lu, G.~l.~Liu, and Q.~j.~Xu,
Phys.\ Rev.\ D {\bf 64}, 095004 (2001).

\bibitem{Lu:2003yr}
G.~r.~Lu, F.~r.~Yin, X.~l.~Wang, and L.~d.~Wan,
Phys.\ Rev.\ D {\bf 68}, 015002 (2003).


\bibitem{Diaz-Cruz:1999ab}
J.~L.~Diaz-Cruz, M.~A.~Perez, G.~Tavares-Velasco, and J.~J.~Toscano,
Phys.\ Rev.\ D {\bf 60}, 115014 (1999).

\bibitem{Cordero-Cid:2004hk}
S.~Bar-Shalom, G.~Eilam, and A.~Soni,
Phys.\ Rev.\ D {\bf 60}, 035007 (1999);
E.~O.~Iltan and I.~Turan,
Phys.\ Rev.\ D {\bf 67}, 015004 (2003);
A.~Cordero-Cid, J.~M.~Hernandez, G.~Tavares-Velasco, and
J.~J.~Toscano,
arXiv:hep-ph/0411188.

\bibitem{Cheng:1987rs}
T.~P.~Cheng and M.~Sher,
Phys.\ Rev.\ D {\bf 35}, 3484 (1987).

\bibitem{Antaramian:1992ya}
A.~Antaramian, L.~J.~Hall, and A.~Rasin,
Phys.\ Rev.\ Lett.\  {\bf 69}, 1871 (1992).

\bibitem{Hall:1993ca}
L.~J.~Hall and S.~Weinberg,
Phys.\ Rev.\ D {\bf 48}, 979 (1993).

\bibitem{Luke:1993cy}
M.~E.~Luke and M.~J.~Savage,
Phys.\ Lett.\ B {\bf 307}, 387 (1993).

\bibitem{Glashow:1976nt}
S.~L.~Glashow and S.~Weinberg,
Phys.\ Rev.\ D {\bf 15}, 1958 (1977).

\bibitem{Passarino:1978jh}
G.~Passarino and M.~J.~G.~Veltman,
Nucl.\ Phys.\ B {\bf 160}, 151 (1979).

\bibitem{Eidelman:2004wy}
S.~Eidelman {\it et al.},
Phys.\ Lett.\ B {\bf 592}, 1 (2004).

\bibitem{vanOldenborgh:1990yc}
G.~J.~van Oldenborgh,
Comput.\ Phys.\ Commun.\  {\bf 66}, 1 (1991).


\bibitem{Atwood:1996vj}
D.~Atwood, L.~Reina, and A.~Soni,
Phys.\ Rev.\ D {\bf 55}, 3156 (1997);
eConf {\bf C960625}, LTH093 (1996); M.~Sher,
arXiv:hep-ph/9809590; T.~M.~Aliev and E.~O.~Iltan,
Phys.\ Rev.\ D {\bf 58}, 095014 (1998).

\bibitem{Haeri} G. Eilam, B. Haeri, and A. Soni, Phys. Rev. D {\bf 41}, 875
(1991).

\bibitem{Arhrib:2004xu}
A.~Arhrib,
arXiv:hep-ph/0409218.

\bibitem{Bejar:2004rz}
S.~Bejar, F.~Dilme, J.~Guasch, and J.~Sola,
JHEP {\bf 0408}, 018 (2004).

\bibitem{Djouadi:1997yw}
A.~Djouadi, J.~Kalinowski, and M.~Spira,
Comput.\ Phys.\ Commun.\  {\bf 108} (1998) 56.

\bibitem{Mertig:1990an}
R.~Mertig, M.~Bohm, and A.~Denner,
Comput.\ Phys.\ Commun.\  {\bf 64}, 345 (1991).

\end{thebibliography}
\end{document}